# Modification of Bayesian Updating where Continuous Parameters have Differing Relationships with New and Existing Data

Nicholas Lewis (Email: nhlewis@btinternet.com)

*Bath, United Kingdom*

12 August 2013

## Abstract

Bayesian analyses are often performed using so-called noninformative priors, with a view to achieving objective inference about unknown parameters on which available data depends. Noninformative priors depend on the relationship of the data to the parameters over the sample space. Combining Bayesian updating – multiplying an existing posterior density for parameters being estimated by a likelihood function derived from independent new data that depend on those parameters and renormalizing – with use of noninformative priors gives rise to inconsistency where existing and new data depend on continuous parameters in different ways. In such cases, noninformative priors for inference from only the existing and from only the new data would differ, so Bayesian updating would give different final posterior densities depending on which set of data was used to derive an initial posterior and which was used to update that posterior. I propose a revised Bayesian updating method, which resolves this inconsistency by updating the prior as well as the likelihood function, and involves only a single application of Bayes' theorem. The revised method is also applicable where actual prior information as to parameter values exists and inference that objectively reflects the existing information as well as new data is sought. I demonstrate by numerical testing the probability-matching superiority of the proposed revised updating method, in two cases.

**1. Introduction**

Bayes' theorem, in the continuous parameter case, formally relates the (posterior) estimated conditional probability density $p_{\theta|x}(\theta|x)$ for a parameter vector $\theta$ to the conditional probability density $p_{x|\theta}(x|\theta)$ for a set of observed variables, or sufficient statistics derived from them, (data) $x$, expressed as a function of $\theta$, and a prior probability distribution (prior), with density $p_\theta(\theta)$, for the parameter vector:

$$p_{\theta|x}(\theta|x) = \frac{p_{x|\theta}(x|\theta) p_\theta(\theta)}{\int p_{x|\theta}(x|\theta) p_\theta(\theta) d\theta} \qquad (1.1)$$

The denominator may alternatively be expressed as $p_x(\boldsymbol{x})$. Where, as here, $p_{x|\theta}(\boldsymbol{x}|\boldsymbol{\theta})$ is considered as a function of $\boldsymbol{\theta}$, at fixed data sample values, it is known as the likelihood (function), $L_x(\boldsymbol{\theta})$. However, it remains a (joint) probability density for $\boldsymbol{x}$, not for $\boldsymbol{\theta}$. (In a case where the data are discrete, $p_{x|\theta}(\boldsymbol{x}|\boldsymbol{\theta})$ represents a probability not a probability density.) The denominator is independent of $\boldsymbol{\theta}$, and can be viewed as a normalising factor that makes the posterior probability sum to unity. Bayes' theorem can be used as a proportionality relationship with this factor omitted, with normalisation carried out at the final stage of the analysis:

$$p_{\theta|x}(\boldsymbol{\theta}|\boldsymbol{x}) \propto p_{x|\theta}(\boldsymbol{x}|\boldsymbol{\theta}) p_\theta(\boldsymbol{\theta}) \qquad (1.2)$$

Where data, through the likelihood function, dominates the shape of the posterior density, the form of the prior distribution matters little. But where that is not the case, due to the data being imprecise – often arising from small sample sizes – the prior may have a major influence on parameter inference. The issue of prior selection has generally been considered in the case where there is negligible prior information as to parameter values, or where it is desired to analyse the data on that basis – as is usual for scientific experiments. Bayes' theorem should not be regarded as requiring that $p_\theta(\boldsymbol{\theta})$ represent an actual estimated unconditional probability density for $\boldsymbol{\theta}$. From the "objective" Bayesian standpoint, the aim is rather to determine a prior that is "noninformative" in the sense of not conveying any information about parameter values, but which rather "lets the data speak for themselves". That is, a prior calculated to achieve inference that reflects to the greatest extent possible even very weak data. A noninformative prior does not represent any particular beliefs as to parameter values and is not interpretable in probabilistic terms (Bernardo and Smith, 1994). Rather, it should be viewed as constituting a mathematical prior or weighting function (Fraser et al., 2011). The weighting provided by $p_\theta(\boldsymbol{\theta})$ can be seen as counteracting the effect of the chosen parameterization, in the light of the likelihood function, and will reflect how, under the chosen parameterization, the average informativeness of the data about parameter values varies therewith.

Although subjective Bayesians may be unconcerned about such matters, the performance of noninformative priors is often judged by their probability-matching: how closely the resulting posterior probabilities agree with repeated-sampling coverage probabilities (Kass and Wassermann, 1996). Indeed, Berger and Bernardo (1992) refer to the commonly used safeguard of frequentist evaluation of the performance of noninformative priors in repeated use, as being historically the most effective approach to discriminating among possible noninformative priors. It is generally, albeit not universally, accepted that an acceptable noninformative prior should – at least asymptotically – lead, faster than the general sample-size driven $O(n^{-1/2})$ convergence rate, to frequentist validity of Bayesian credible sets. That is to say, it should be a probability matching prior (see, e.g., Rubin, 1984; Berger and Wolpert, 1984, p. 73; Tibshirani, 1989; Kass and Wasserman, 1996; Mukerejee and Reid, 1999; Little, 2011).

Noninformative priors present difficult problems and have given rise to much debate. In most cases it appears impossible – even allowing priors to be functions of the data as well as of the parameters – to formulate priors that are completely noninformative, in the sense of achieving inference that is exactly probability matching. However, there are recognised methods for seeking priors that may be expected to have a minimal effect, relative to the data, on parameter inference. Jeffreys pioneered the development of noninformative priors for continuous parameters, leading both to the original (nonlocation) Jeffreys' prior, referred to in this paper simply as Jeffreys' prior, and its location parameter variant. Jeffreys' prior is the square root of the determinant of the expected Fisher information matrix $h(\boldsymbol{\theta})$:

$$h(\boldsymbol{\theta}) = -\int p_{x|\boldsymbol{\theta}}(\boldsymbol{x}|\boldsymbol{\theta}) \frac{\partial^2 \log p_{x|\boldsymbol{\theta}}(\boldsymbol{x}|\boldsymbol{\theta})}{\partial \boldsymbol{\theta}^2} d\boldsymbol{x} \tag{1.3}$$

A key feature of Jeffreys' prior is that it corresponds to a natural volume element of a Riemannian metric. Such volume elements are invariant under both data transformations and reparameterisation and generate locally uniform measures on manifolds, in the sense that equal mass is assigned to regions having equal volumes (Kass, 1989; Kass and Wassermann, 1996). Put another way, Jeffreys' prior is a uniform distribution taking account of the topology (Ghosh et al, 2006). As a result, Jeffreys' prior is invariant: for the parameters jointly, posterior inference is consistent (i.e., identical when expressed in the same parameterisation) under changes of the data and/or parameter variables. Jeffreys' prior can be viewed as a density conversion factor between natural volume elements in parameter space and corresponding volumes in data space, multiplication by which converts the likelihood function from a probability density in data space to a probability density in parameter space.

In the univariate parameter case it is known (Hartigan, 1965) that when Jeffreys' prior, $|h(\boldsymbol{\theta})|^{1/2}$, is used ($|\ldots|$ denoting a matrix determinant), one-sided Bayesian credible intervals are, asymptotically, closer in probability to confidence intervals for the same parameter

regions – matching to $O(n^{-1})$ (Welch, 1965) – than when any other prior is used. If and only if $\theta$, or some monotonic transformation of $\theta$, is a location parameter with respect to the data or some transformation thereof, and the model admits a real-valued sufficient statistic, then one-sided credible interval and confidence interval probabilities match exactly (Lindley, 1958). Where the aim is objective inference in the absence of prior information, it is now generally accepted that, as Bernardo and Smith (1994) put it: "In one-dimensional continuous regular problems, Jeffreys' prior is appropriate".

Apart from Jeffreys' priors, probably the best known approach to formulating noninformative priors is Bernardo and Berger's reference analysis (Bernardo, 1979; Berger and Bernardo, 1992), which involves maximising the (expected value of) missing information about the parameters of interest. In a univariate parameter setting, these two approaches coincide provided that – as is usually the case – the asymptotic posterior distribution for the parameter is normal (Bernardo and Smith, 1994, p. 314). In the multivariate case, Jeffreys' prior remains the reference prior for inference about all parameters jointly, taken as a single group (Robert, 2007). However, satisfactory marginal inference about a subset of parameter(s) of interest is more reliably achieved by reference analysis than by using Jeffreys' prior. The least informative, "reference" prior may vary depending on what priority ordering for marginal inference is given to individual components or subsets of the parameter vector. Berger and Bernardo's ordered reference priors for marginal inference are, like Jeffreys' prior, computed from the Fisher information matrix. Additionally, Peers (1965) gave a sufficient condition for a prior to be approximately probability-matching, which is also expressed in terms of the Fisher information matrix. Accordingly, in general Bayesian parameter inference based on a noninformative prior, with the aim of achieving approximate probability-matching, requires the availability of the Fisher information matrix pertaining to all the data on which that inference is based, and can typically be computed from that matrix.

Some Bayesians object to the sample space dependence of Jeffreys' and other noninformative priors – their dependence on the experiment involved, through its influence on the likelihood function and/or the sampling procedure. That dependence violates the so-called (strong) likelihood principle, supposedly proved by Birnbaum (1962). However, the likelihood principle – which is incompatible with frequentist statistical methodology – is doubted by many statisticians, at least insofar as it is applied to statistical inference. See, e.g., Evans et al. (1986) and Mayo (2010). More recently, Evans (2013) has shown that the theorem Birnbaum proved is incorrectly stated, and Mayo (2013) shows how data may violate the likelihood principle while upholding both the weak conditionality and the sufficiency principles.

## 2. Bayesian updating

Bayes' theorem is often used to carry out sequential analysis and revision of beliefs, using the posterior derived from the application of Bayes' theorem to the results of one experiment, A, as the prior for inference based on the results of a second experiment, B. A new posterior is derived as the renormalized product of the old posterior (used as a prior) and the likelihood function for the second experiment:

$$p_{\theta|x_A x_B}(\theta | x_A x_B) \propto p_{x_B|\theta}(x_B | \theta) [p_{x_A|\theta}(x_A | \theta) p_\theta(\theta)] \qquad (2.1)$$

This procedure is termed "Bayesian updating". It is commonly considered, even by objective Bayesians, to be a valid means of inference irrespective of the whether or not the new data comes from a similar experiment to that involved in the original application of Bayes' theorem (see, e.g., Zellner, 1971, p17; Box and Tiao, 1973, p11; Jaynes, 2003, p259; or Sivia, 2006, p20). Bayesian updating respects the likelihood principle, since inference depends on the evidence provided by experiment B only through its likelihood function.

This paper shows, through both mathematical analysis and numerical testing, that the standard Bayesian updating procedure will generally not result in posterior probabilities that achieve optimal probability matching, when, because of differing non-linear data-parameter relationships or otherwise, Fisher information matrices differ (up to proportionality) between experiments. It also shows how to revise Bayesian updating to overcome this problem.

I adopt an objective Bayesian position and proceed at present on the basis that no actual prior information as to parameter values exists, so that using a noninformative prior is appropriate. It is first demonstrated that standard Bayesian updating can produce inconsistent results when noninformative priors are used, a fact noted by Hartigan (1964) and by Kass and Wasserman (1996). An (expected) Fisher information based approach is used. Jeffreys' prior is used to motivate the analysis, but since the key point is that Fisher information – which is generally central to noninformative prior formulations – does not update in the same way as likelihood, the findings are not limited to where Jeffreys' prior is used.

For ease of analysis, certain simplifying assumptions are made. It is assumed that the probability model is known, dependent on a vector of unknown parameters taking continuous values, and to be such that the asymptotic joint parameter posterior distribution is (multivariate) normal, as is usually the case. Likelihood functions are assumed to be smooth well-behaved functions, twice differentiable at all points. The deficiency identified in standard Bayesian updating, and the proposed revisions thereto, are in point however many parameters and experiments (each giving rise to one or more data realisations) there are. However, for ease of exposition – and also because with a univariate parameter Jeffreys' prior is known to provide the best probability matching – only the case with a univariate real-valued parameter $\theta$ and two experiments, A and B, is analysed in detail.

Suppose that experiments A and B give rise respectively to data $x_A$ and $x_B$, with probability elements $p_{x_A|\theta}(x_A|\theta)\big|_\theta dx_A$ and $p_{x_B|\theta}(x_B|\theta)\big|_\theta dx_B$ (if the data are discrete $dx_A$ and $dx_B$ are inapplicable). It is assumed that $x_A$ and $x_B$ are conditionally independent given $\theta$, so that (dropping the subscripts on $p$) if the two experiments are viewed as a single composite experiment C giving rise to a data sample $x_C = \{x_A, x_B\}$,

$$p(x_C|\theta)\big|_\theta dx_C = p(x_A|\theta)\big|_\theta \, p(x_B|\theta)\big|_\theta dx_A dx_B$$

where $dx_C = dx_A dx_B$. The likelihoods $L_A(\theta) = p(x_A|\theta)\big|_{x_A}$ and $L_B(\theta) = p(x_B|\theta)\big|_{x_B}$ have expected Fisher information (respectively $h_A$ and $h_B$) defined by:

$$h_A(\theta) = -\int p(x_A|\theta) \frac{\partial^2 \log L_A}{\partial \theta^2} dx_A \qquad (2.2)$$

$$h_B(\theta) = -\int p(x_B|\theta) \frac{\partial^2 \log L_B}{\partial \theta^2} dx_B \qquad (2.3)$$

Since the appropriate noninformative prior is Jeffreys' prior, the inconsistency arising from Bayesian updating when the two experiments involved, A and B, embody differing expected Fisher information can be demonstrated simply. It suffices to work with proportionality rather than equality. Suppose that data from experiment A is analysed first, and Bayes' theorem is used with the Jeffreys' prior for that experiment to obtain a posterior density for $\theta$:

$$p(\theta|x_A) \propto L_A(\theta)\big|h_A(\theta)\big|^{1/2} \qquad (2.4)$$

Bayes' theorem is then used again to update the thus-obtained posterior with the likelihood from experiment B's data. That gives a posterior density for $\theta$ reflecting information from both experiments' data:

$$p(\theta | x_A x_B) \propto L_B(\theta) p(\theta | x_A) \propto L_B(\theta) L_A(\theta) |h_A(\theta)|^{1/2} \qquad (2.5)$$

Alternatively, if data from experiment B is analysed first, using Bayes' theorem with the Jeffreys' prior for that experiment one obtains the following posterior density for $\theta$:

$$p(\theta | x_B) \propto L_B(\theta) |h_B(\theta)|^{1/2} \qquad (2.6)$$

Use of Bayes' theorem again to update the thus-obtained posterior with the likelihood from experiment A's data sample also gives a posterior density for $\theta$ reflecting information from both experiments' data:

$$p(\theta | x_B x_A) \propto L_A(\theta) p(\theta | x_B) \propto L_A(\theta) L_B(\theta) |h_B(\theta)|^{1/2} \qquad (2.7)$$

Since multiplication of likelihoods is commutative, it is evident that from (2.5) and (2.7) that $p(\theta | x_B x_A) = p(\theta | x_A x_B)$ for all $\theta$ if and only if $|h_B(\theta)| \propto |h_A(\theta)|$. (If the Fisher information matrix determinants, and hence the priors, are proportional, normalisation to unity of total probability will ensure equality, not just proportionality, of the posterior densities.)

Where the probability models for the two experiments A and B have different forms, typically reflecting different types of data-parameter relationships or data error distributions, their expected Fisher information matrices will not generally have proportionate determinants, so the final posterior density for $\theta$ will depend on the order in which the experiments are analysed. This inconsistency indicates that something is wrong. The order in which evidence from two experiments is analysed and combined should not affect inference. Kass and Wassermann (1996) expressed concern that when using noninformative priors Bayesian updating can generate a different posterior depending on the order that data is processed, but did not focus on the possibility that the problem lay with Bayesian updating.

Although the foregoing arguments have been set out for a univariate parameter, they generalise immediately to the multivariate case, with the priors, likelihoods and posteriors then being joint densities.

## 3. Modifying Bayesian updating

The appropriate noninformative prior to use for parameter inference based on data from both experiment A and experiment B is readily found by considering them as a single composite experiment, C. Assuming independence (conditional on $\theta$) of the two experiments' data, the likelihood function for the composite experiment is given by multiplying those for experiments A and B:

$$L_C(\theta) = p(\boldsymbol{x}_C | \theta)\big|_{\boldsymbol{x}_C} = p(\{\boldsymbol{x}_A, \boldsymbol{x}_B\} | \theta)\big|_{\{\boldsymbol{x}_A, \boldsymbol{x}_B\}} = p(\boldsymbol{x}_A | \theta)\big|_{\boldsymbol{x}_A} p(\boldsymbol{x}_B | \theta)\big|_{\boldsymbol{x}_B} = L_A(\theta) L_B(\theta) \qquad (3.1)$$

One can likewise express the expected Fisher information for the composite experiment:

$$h_C(\theta) = -\int\int [p(\boldsymbol{x}_A | \theta) p(\boldsymbol{x}_B | \theta)] \frac{\partial^2 \log[L_A L_B]}{\partial \theta^2} d\boldsymbol{x}_A d\boldsymbol{x}_B \qquad (3.2)$$

(where the fact that $d\boldsymbol{x}_C = d\boldsymbol{x}_A d\boldsymbol{x}_B$ has been used) in terms of that for the separate experiments. Using the additivity of logarithms and separating the integrals using the conditional independence of $\boldsymbol{x}_A$ and $\boldsymbol{x}_B$ gives:

$$h_C(\theta) = -\int p(\boldsymbol{x}_B | \theta) d\boldsymbol{x}_B \int p(\boldsymbol{x}_A | \theta) \frac{\partial^2 \log L_A}{\partial \theta^2} d\boldsymbol{x}_A - \int p(\boldsymbol{x}_A | \theta) d\boldsymbol{x}_A \int p(\boldsymbol{x}_B | \theta) \frac{\partial^2 \log L_B}{\partial \theta^2} d\boldsymbol{x}_B \qquad (3.3)$$

Since the first integral in each term evaluates to unity and the second integral is the Fisher information for the individual experiment, it follows that the Fisher information is additive:

$$h_C(\theta) = h_A(\theta) + h_B(\theta) \qquad (3.4)$$

Applying Bayes' theorem using Jeffreys' prior, $|h_C(\theta)|^{1/2}$, and the likelihood function for the composite experiment C:

$$p(\theta | \boldsymbol{x}_C) \propto L_C(\theta) |h_C(\theta)|^{1/2} \qquad (3.5)$$

which, expressed in terms relating to the underlying separate experiments, gives:

$$p(\theta | \boldsymbol{x}_A, \boldsymbol{x}_B) \propto [L_A(\theta) L_B(\theta)] |h_A(\theta) + h_B(\theta)|^{1/2} \qquad (3.6)$$

Since addition and multiplication are commutative, this posterior is, as it logically should be, unaffected by the order in which the experimental data is analysed. In order, where Jeffreys' prior is used, for Bayesian updating to produce inference that is consistent with that from one-step Bayesian inference for the combined experiment, it must take the revised form given in (3.6). That is to say, separate terms for the likelihood and the Fisher information matrix are required, updated respectively by multiplication and by addition. In geometric terms, the Fisher information term can be seen as building up a model parameter space metric, which becomes denser as more information is obtained, the volume density of the space being the square root of the determinant of the metric (Mosegaard and Tarantola, 2002).

The proposed revised basis of updating is identical if some other noninformative prior computed from the Fisher information matrix $h$, such as a reference prior, is to be used in place of Jeffreys' prior. Suppose that prior takes the form $f(h(\theta))$, where $f$ can be any function, including one implicitly specified by an algorithm (as in reference analysis). In that case, (3.6) is replaced by the more general form:

$$p(\theta | x_A, x_B) \propto [L_A(\theta) L_B(\theta)] f(h_A(\theta) + h_B(\theta)) \qquad (3.7)$$

In order to implement the revised updating procedure, it is necessary to maintain separate records of the combined (cumulatively multiplied) likelihood functions and of the combined (cumulatively added) Fisher information matrices from all experiments carried out to date, rather than just a record of the posterior density. Provided that the intended noninformative prior depends only on the Fisher information matrix, doing so enables correct revision of the posterior density as new data becomes available, whether from a new experiment or from one of the already incorporated experiments. It can readily be seen that, no matter how many experiments and sets of data are involved, the posterior density derived from the revised Bayesian updating method will be identical to that which would arise were all the data to be aggregated and analysed simultaneously, as if arising from a single composite experiment.

At each updating using the revised procedure, first the cumulative sum of the expected Fisher information matrices and the cumulative product of the likelihood functions are updated to reflect the new experiment and data therefrom. A noninformative prior is then computed from the updated Fisher information matrix, and multiplied by the updated product of the likelihood functions. The result is normalised to give the updated joint posterior density for the parameter vector. Marginal posteriors can then be produced in the usual way by integrating out nuisance parameters. In a univariate parameter setting, or for inference as to all parameters jointly in a multiparameter case, the appropriate noninformative prior to use would be Jeffreys' prior, calculated directly from the determinant of the updated Fisher information. Where in a multivariate parameter case marginal inference is to be made about a subset of one or more parameters of interest, ordered reference priors could be computed from the full Fisher information matrix (Bernardo and Smith, 1994, Proposition 5.30). Alternatively, an approximate probability matching prior might be constructed, for instance using the method set out in Sweeting (2005). However, should a data-dependent noninformative prior be used, then to ensure consistency Bayesian updating would require maintaining a record of the data from all the experiments in addition to the cumulative product of their likelihood functions and sum of their Fisher information matrices.

If the Fisher information matrices for the two experiments are identical up to proportionality then so will be the Fisher information for the combined experiments: $h_C(\theta) = h_A(\theta) + h_B(\theta) = h_A(\theta) + ch_A(\theta) = kh_A(\theta)$. In that event standard Bayesian updating needs no modification when using a noninformative prior that is a function of the Fisher information matrix, since (3.7) reduces to:

$$p(\theta | x_A, x_B) \propto [L_A(\theta) L_B(\theta)] f(h_A(\theta)) \propto L_B(\theta) p(\theta | x_A) \qquad (3.8)$$

An example is where the second experiment consists of obtaining an additional sample under the same setup as in the first experiment.

## 4. Incorporation of probabilistic prior information

Although the foregoing analysis has been undertaken in the context where there is either no existing probabilistic information as to parameter values before the experiments under consideration, or such information is to be ignored, it applies equally where such information exists and is to be incorporated. It is assumed for simplicity that such information is, conditional on the parameter value(s), independent of the experimental data involved. When such actual prior information exists, one can suppose that it is equivalent to a notional observation $y$ with a certain probability density $p(y|\theta)$, from which a posterior density of $\theta$ given $y$ has been calculated using Bayes' theorem with a noninformative prior. Doing so was proposed by Hartigan (1965), with the thus computed posterior density being employed as the prior density representing the existing information in a further application of Bayes' theorem (a Bayesian update) in combination with the likelihood function for the actual observation(s).

Indeed, if the existing information is not treated in the foregoing way, as a likelihood function pertaining to a notional observation multiplied by an appropriate noninformative prior, then in general the posterior density cannot correctly reflect the combination of the existing information with that from the experimental data. That must be so since the issues with Bayesian updating that have been identified apply equally where an intentionally informative prior density is updated by the probability density of actual observation(s) from a single experiment, as they do in the case with two experiments but no incorporation of existing information. That informative probabilistic prior information cannot in general be correctly incorporated by simply using it as a prior density and applying Bayes' theorem can be seen by considering the case where, in the two experiments case, inference from one of the experiments on its own gives a posterior density identical to the informative prior density and the other experiment is the same as the single experiment in the informative prior information case.

Accordingly, adopting Hartigan's viewpoint is necessary for the correct incorporation of actual prior information. However, the proposed revised updating procedure should be used rather than simply applying Bayes' theorem twice as proposed by Hartigan. It is therefore generally necessary to calculate the Fisher information pertaining to the probability density for the notional observation that represents actual prior information. An exception is when the Fisher information matrix relating to the notional observation can be seen to be proportional to the Fisher information matrix relating to the actual observation(s).

## 5. Numerical testing

Probability matching accuracy of parameter inference resulting from use of the standard and the revised Bayesian updating methods is now investigated. Two cases are examined, each involving two independent experiments and a univariate parameter for which asymptotic normality applies. Since the product of the likelihood functions for the individual experiments is identical for both types of updating, differences in posterior inferences arise entirely from different priors. Jeffreys' prior is used for all inference, since it is known to produce the best probability matching here. Testing involves generating a large number of random data samples according to the assumed data distributions, with one or more known parameter values. Where upper and lower bounds have to be imposed for computational reasons, they are set sufficiently wide for all likelihood functions to have negligible weight beyond the bounds. Posterior PDFs are normalised to integrate to unity between the bounds used. The posterior probability of the Bayesian one-sided credible prediction intervals – derived from the marginal posterior cumulative density functions (CDFs), and invariant under reparameterisation – is compared with the corresponding frequentist coverage probability. Specifically, the proportions of cases in which the actual parameter value falls below the parameter values at which the Bayesian posterior CDFs cross each integral cumulative probability percentage point is computed. For a completely uninformative prior, and if Bayesian inference exactly reflected long-run repeated-sampling probabilities, the relationship of those proportions with the CDF percentage points should be linear with a unit slope and zero intercept, subject to random deviations due to the number of samples being finite.

Lindley (1958) showed that, whatever prior is used, Bayes one-sided credible regions and frequentist confidence intervals cannot always coincide unless a location parameter is involved. A location parameter is involved if, for some monotonic transformations of the observed and parameter variables, the probability distribution of the transformed observation depends only on the difference between the two transformed variables (e.g., Gelman et al, 2004, p.64). There is an illuminating discussion of Bayesian inference in Fraser and Reid (2011), where it is shown how, where there is more than one data variable, curvature (a non-linear relationship) in the model or in the parameter of interest leads to Bayesian credible regions giving results that do not match those from repeated-sampling. Therefore, as the two-experiment examples considered here involve curvature in the parameter–variable relationship, one cannot expect them to give exact probability matching even with the most uninformative prior possible. Nevertheless, with a correctly formulated noninformative prior, resulting from use of the revised Bayesian updating method, probability matching should be superior to that with an incorrectly formulated prior.

*Example 1: Inference where data likelihoods depend on respectively a parameter and its cube*

Consider experiment A where the datum $x_A$ is related to the parameter $\theta$ by the (location parameter) probability model:

$$x_A = \theta + e_A \quad \text{with } e_A \sim N(0, \sigma_A) \tag{5.1}$$

which implies that the likelihood function is:

$$p(x_A | \theta) = (\sqrt{2\pi}\sigma_A)^{-1} \exp\left(-(x_A - \theta)^2 / 2\sigma_A^2\right) \tag{5.2}$$

In such a case the Jeffreys' prior for $\theta$ is uniform ($p(\theta)$ = constant), is the reference prior and is completely noninformative. Therefore, in the absence of any other relevant information about $\theta$ before analysing experiment A, we can apply (1.2) with $p(\theta)$ a constant, obtaining:

$$p(\theta|x_A) \propto p(x_A|\theta) \tag{5.3}$$

Now consider experiment B, where the datum $x_B$ is related to the same parameter $\theta$ by the probability model:

$$x_B = \theta^3 + e_B \quad \text{with } e \sim N(0, \sigma_B) \tag{5.4}$$

which implies that the likelihood function is:

$$p(x_B | \theta) = (\sqrt{2\pi}\sigma_B)^{-1} \exp\left(-(x_B - \theta^3)^2 / 2\sigma_B^2\right) \tag{5.5}$$

In this case, since Jeffreys' prior is invariant under one-to-one parameter transformations and Jeffreys' prior for $\theta^3$ would be uniform (c/f experiment A), under experiment B Jeffreys' prior (and the reference prior) for $\theta$ is proportional to $\theta^2$. (Conversion of a PDF for $\theta^3$ into a PDF for $\theta$ using the standard Jacobian formula for transforming a probability density on a change of variables (e.g., Sivia, 2006, p.69) involves multiplication by $|d\theta^3/d\theta| = 3\theta^2$.) Applying equation (1.2) using the results of experiment B alone, we therefore have:

$$p(\theta|x_B) \propto p(x_B|\theta)\,\theta^2 \tag{5.6}$$

In physical terms, $\theta$ might be the radius of a sphere, estimated in experiment A by measuring its circumference and in experiment B by measuring its volume, with in each case the measurement being subject a normally distributed random error with known standard deviation.

Applying the standard Bayesian updating formula (2.1) to incorporate the results of experiment B where Bayesian inference has already been undertaken using the results of experiment A, the posterior reflecting the evidence from both experiments is:

$$p(\theta | x_A x_B) \propto p(x_B | \theta)\, p(\theta | x_A) \tag{5.7}$$

Substituting from (5.3) we obtain:

$$p(\theta | x_A x_B) \propto p(x_B|\theta)\, p(x_A|\theta) \tag{5.8}$$

It would be equally acceptable to analyse the results of experiment B first and then to update the thereby-derived posterior for $\theta$ with the results of experiment A, using:

$$p(\theta | x_B x_A) \propto p(x_A | \theta)\, p(\theta | x_B) \tag{5.9}$$

Substituting from (5.6) we obtain:

$$p(\theta | x_B x_A) \propto p(x_A | \theta) p(x_B | \theta) \theta^2 \qquad (5.10)$$

Equations (5.8) and (5.10) each represent a posterior PDF for $\theta$ that reflects the evidence from both experiments in accordance with Bayesian updating, using a noninformative reference prior for the first experiment analysed. But they differ by a factor of $\theta^2$.

In fact, neither equation (5.8) or (5.10) incorporates a prior that is noninformative for analysing the combined results of the two experiments. And although with enough data from either experiment the two posteriors will converge on the same (Gaussian) distribution, with weak data the posteriors produced by (5.8) and (5.10) will be very different. As shown in Appendix 1, Jeffreys' prior (and hence the reference prior) for inference based on the combined results of experiments A and B is neither uniform nor proportional to $\theta^2$. Instead, it is (per A1.9):

$$\pi(\theta) = (\sigma_A^{-2} + \sigma_B^{-2} 9\theta^4)^{1/2} \qquad (5.11)$$

Panel (a) of Figure 1 shows that for experiments A and B on their own, using their respective uniform and $\theta^2$ Jeffreys'/reference priors, with the posteriors generated in accordance with equations (5.3) and (5.6) respectively, probability matching is very accurate in both cases. Panel (b) of Figure 1 shows that for experiments A and B combined, probability matching is substantially inaccurate using the Jeffreys'/reference prior for either experiment (in accordance with the standard Bayesian updating equations (5.8) and (5.10) respectively), but is far more accurate when the Jeffreys'/reference prior (per equation (5.11)) for the combined experiments is used, in accordance with the revised Bayesian updating method.

When using a uniform prior, corresponding to first applying Bayes' theorem to experiment A using the reference prior, and then updating the posterior with the results of experiment B, rather than the true parameter value lying below the 5th or above the 95th percentage point of the posterior CDF in only 5% of instances for both cases, it did so in respectively 2.0% and 8.7% of instances. Using a $\theta^2$ prior, corresponding to first applying Bayes' theorem to experiment B using reference prior, and then updating the posterior with the results of experiment A, the corresponding figures were 7.0% and 2.4%. Using the reference prior for the combined experiments, these figures were respectively 5.7% and 5.6%, both close to a perfect match.

*Example 2: Inference for Bernoulli trials*

Consider now the canonical Example 9 in Berger and Wolpert (1984), of two experiments where either the number of zeros *z* for a fixed sample size *n* of Bernoulli i.i.d. random variables is counted (experiment A, giving a binomial distribution), or the number of observations *y* is counted until a fixed number *r* of zeros (failures) occur (experiment B, giving a negative binomial distribution). Here the parameter is continuous but the data are discrete. Box and Tiao (1973, p.45) noted that the Jeffreys' priors for these two experiments differ, and concluded that Bayesian inference from them should differ even when the likelihood functions are identical, which they are when $z = r$ and $y = n$. The Jeffreys' priors they derived for the binomial and negative binomial distribution cases, rederived as (2.8) and (2.16) in Appendix 2, were respectively:

$$\pi(\theta) \propto \theta^{-1/2}(1-\theta)^{-1/2} \quad (5.12)$$

and

$$\pi(\theta) \propto \theta^{-1}(1-\theta)^{-1/2} \quad (5.13)$$

where $\theta$ is the probability of the outcome being counted (here, failures) occurring in any trial.

Berger and Wolpert (1984) pointed out that the analysis by Box and Tiao violated the (strong) likelihood principle. However, if one regards a noninformative prior as reflecting how informative the data can be expected to be about different parameter values, one would expect a noninformative prior to increase more rapidly as $\theta$ decreases in the negative binomial case. That is because the lower $\theta$ is, the greater the expected number of trials to obtain a fixed number of the outcomes being counted, and hence the more information the data will contain about $\theta$ compared to the binomial case, where the number of trials is fixed.

Consider now inference based on the combined results of two independent experiments, one of each type. As per equation (A2.25), Jeffreys' prior for the combined experiments, used in the revised method of Bayesian updating, has the form:

$$\pi(\theta) \propto (r+n\theta)^{1/2}\theta^{-1}(1-\theta)^{-1/2} \quad (5.14)$$

The difference in the two priors, and hence of inference in a small sample case, is material only if the probability of the outcomes being counted is small. The case where the probability of failure varies between 0.01 and 0.11, with 40 trials and 2 failures (zeros), is therefore considered. Panel (a) of Figure 2 shows that for experiments A and B on their own, using their respective Jeffreys' priors, probability matching is in each such case much better than where the Jeffreys' prior appropriate for the other experiment is used. This supports Box and Tiao's analysis and provides evidence against the likelihood principle.

Panel (b) of Figure 2 shows that for experiments A and B combined, probability matching is much more accurate when Jeffreys' prior for the combined experiments is used, as under the revised method of Bayesian updating, than when Jeffreys' prior for either separate experiment is used (according to which is analysed first), as would be the case under standard Bayesian updating.

## 6. Discussion

In the continuous parameter case, from an objective viewpoint standard Bayesian updating is unsatisfactory unless all the inferential data involved comes either from the same experiment or from experiments that all imply the same noninformative prior. In other cases Bayesian updating needs to be modified so that Bayes' theorem is applied once only, to the multiplicatively combined likelihoods from the data from all the experiments (assuming conditionally independence of the experimental data), using whatever is considered a suitable noninformative prior for inference about the parameter(s) of interest from the combined data (e.g., Jeffreys' prior or a targeted reference prior). Except where a data-dependent prior is to be used, such a noninformative prior would generally be computed from the expected Fisher information for the combined experiments, being the sum of that for the individual experiments.

The problem with standard Bayesian updating in the continuous parameter case arises because the likelihood function for observed data does not contain all the information provided by an experiment, and nor – even when it is exactly probability matching – does a Bayesian posterior density preserve all the information necessary for correct inference when dissimilar additional data is subsequently introduced. Since Fisher information for independent experiments combines additively, the proposed revision to standard Bayesian updating can be accommodated by preserving the cumulative multiplicatively-combined (joint) likelihood function and the cumulative additively-combined Fisher information matrix in place of just the (joint) posterior density for the parameters, provided that a data-dependent prior is not used.

Numerical testing results indicate superiority of the proposed revised updating approach over standard Bayesian updating, in terms of probability matching under repeated testing. These results support the view that the inconsistency problem mentioned in Kass and Wasserman (1996) reflects a deficiency in the standard Bayesian updating formula, not a problem with noninformative priors.

As discussed in Section 4, the identified shortcomings of standard Bayesian updating also potentially arise, and can likewise be overcome, in the case where there is only one set of experimental data but probabilistic prior information exists as to parameter values. That is because an informative prior density can be viewed as arising from the application of Bayes' theorem to a likelihood function for a notional observational that reflects the actual prior information, using a noninformative prior derived from the Fisher information for that likelihood function. That noninformative prior may differ from the one implied by the Fisher information for the actual data likelihood function. If so, the revised Bayesian updating method should be used in order to correctly combine prior information and experiment data for inferential purposes.

**Appendix 1**: **Derivation of the Jeffreys'/reference prior for two experiments involving a location parameter model but with data relating to different functions of the parameter**

Consider two experiments, A and B, providing information on parameter $\theta$. In experiment A, function $f_A(\theta)$ is measured inaccurately by variable $x_A$, such that $p(x_A\text{-}\theta|\theta) = f(e)$, where e = $(x_A - f_A(\theta))/\sigma_A$ represents standardised measurement errors in $x_A$, known to have a Gaussian distribution with zero mean and standard deviation $\sigma_A$. In experiment B, function $f_B(\theta)$ is measured inaccurately by variable $x_B$, such that $p(x_B\text{-}f_B(\theta)|\theta) = f(e)$, where e = $(x_B - f_B(\theta))/\sigma_B$ represents standardised measurement errors in $x_B$, known to have a zero mean Gaussian distribution with standard deviation $\sigma_B$. The aim is to find a noninformative reference prior which permits objective inference as to the posterior distribution of $\theta$ based on analysing the combined results of experiments A and B. Since the regularity conditions required for the posterior for $\theta$ to be asymptotically normal apply, the noninformative reference prior $\pi(\theta)$ is Jeffreys' prior, the square root of the expected Fisher information $h(\theta)$ (Bernardo and Smith, 1994, pp 287-292 and 314).

Defining the two likelihood functions involved as:

$$p(x_A | \theta) = (\sqrt{2\pi}\sigma_A)^{-1} \exp\left(-(x_A - f_A(\theta))^2 / 2\sigma_A^2\right) \quad (A1.1)$$

$$p(x_B | \theta) = (\sqrt{2\pi}\sigma_B)^{-1} \exp\left(-(x_B - f_B(\theta))^2 / 2\sigma_B^2\right) \quad (A1.2)$$

The expected Fisher information is:

$$h(\theta) = -\int_{x_A}\int_{x_B} p(x_A | \theta) p(x_B | \theta) \frac{\partial^2}{\partial \theta^2} \log\left(p(x_A | \theta) p(x_B | \theta)\right) dx_A dx_B \quad (A1.3)$$

where the integrals are over $(-\infty,\infty)$. We first calculate the derivative:

$$\frac{\partial^2}{\partial \theta^2} \log\left(p(x_A | \theta) p(x_B | \theta)\right) = \frac{\partial^2}{\partial \theta^2}\left(\log(\sqrt{2\pi}\sigma_A)^{-1} - (x_A - f_A(\theta))^2 / 2\sigma_A^2 + \log(\sqrt{2\pi}\sigma_B)^{-1} - (x_B - f_B(\theta))^2 / 2\sigma_B^2\right)$$

$$= -\left(\sigma_A^{-2}(f_A^{'}(\theta)^2 - (x_A - f_A(\theta))f_A^{''}(\theta)) + \sigma_B^{-2}(f_B^{'}(\theta)^2 - (x_B - f_B(\theta))f_B^{''}(\theta))\right) \quad (A1.4)$$

Therefore:

$$h(\theta) = \int_{x_A}\int_{x_B} p(x_A | \theta) p(x_B | \theta)\left(\sigma_A^{-2}(f_A^{'}(\theta)^2 - (x_A - f_A(\theta))f_A^{''}(\theta)) + \sigma_B^{-2}(f_B^{'}(\theta)^2 - (x_B - f_B(\theta))f_B^{''}(\theta))\right) dx_A dx_B$$

$$(A1.5)$$

Since $\int_{x_A} p(x_A|\theta)dx_A = 1;\ \int_{x_B} p(x_B|\theta)dx_B = 1$ :

$$h(\theta) = \int_{x_A} p(x_A|\theta)\sigma_A^{-2}(f_A^{'}(\theta)^2 - (x_A - f_A(\theta))f_A^{''}(\theta))dx_A + \int_{x_B} p(x_B|\theta)\sigma_B^{-2}(f_B^{'}(\theta)^2 - (x_B - f_B(\theta))f_B^{''}(\theta))dx_B$$

Now, $\int_{x_A} p(x_A|\theta)x_A\, dx_A = f_A(\theta)$ and $\int_{x_B} p(x_B|\theta)x_B\, dx_B = f_B(\theta)$ since they are the means of Gaussian distributions with central parameters of respectively $f_A(\theta)$ and $f_B(\theta)$. Therefore the second term in each integral ingrates to zero, giving:

$$h(\theta) = \int_{x_A} p(x_A|\theta)\sigma_A^{-2} f_A^{'}(\theta)^2\, dx_A + \int_{x_B} p(x_B|\theta)\sigma_B^{-2} f_B^{'}(\theta)^2\, dx_B \qquad (A1.6)$$

which reduces to:

$$h(\theta) = \sigma_A^{-2} f_A^{'}(\theta)^2 + \sigma_B^{-2} f_B^{'}(\theta)^2 \qquad (A1.7)$$

from which it follows that the noninformative prior $\pi(\theta) = h(\theta)^{1/2}$ is given by:

$$\pi(\theta) = (\sigma_A^{-2} f_A^{'}(\theta)^2 + \sigma_B^{-2} f_B^{'}(\theta)^2)^{1/2} \qquad (A1.8)$$

In the case of experiments A and B in Example 1 in Section 5, $f_A(\theta) = \theta$ and $f_B(\theta) = \theta^3$, implying that $f'_A(\theta) = 1$ and $f'_B(\theta) = 3\theta^2$. Therefore, by (A1.8) Jeffreys' prior is:

$$\pi(\theta) = (\sigma_A^{-2} + \sigma_B^{-2} 9\theta^4)^{1/2} \qquad (A1.9)$$

**Appendix 2**. **Derivation of the Jeffreys'/reference prior for Bernoulli experiments**

*a) Experiment A: Fixed number of trials – Binomial distribution*

Let the probability of success in any trial be $\theta$, the specified number of trials be $n$, and the number of successes that are achieved be $y$. Then:

$$P(y|\theta,n) = \binom{n}{y}\theta^y(1-\theta)^{n-y} \qquad (A2.1)$$

where $P()$ represents probability, not probability density, since $y$ is a discrete variate.

$$\log P(y|\theta,n) = \log\binom{n}{y} + y\log\theta + (n-y)\log(1-\theta) \qquad (A2.2)$$

$$\frac{\partial^2}{\partial\theta^2}\log P(y|\theta,n) = -y\theta^{-2} - (n-y)(1-\theta)^{-2} \qquad (A2.3)$$

We now compute Fisher's information (matrix) $h(\theta)$, being:

$$h(\theta) = -E\left(P(y|\theta,n)\frac{\partial^2}{\partial\theta^2}\log P(y|\theta,n)\right) = \sum_y \binom{n}{y}\theta^y(1-\theta)^{n-y}(y\theta^{-2}+(n-y)(1-\theta)^{-2}) \quad (A2.4)$$

which since $\sum_y P(y|\theta,n) = 1$ simplifies to:

$$h(\theta) = n(1-\theta)^{-2} + (1-\theta)^{-2}(\theta^{-2} - 2\theta^{-1} + 1 - 1)\left(\sum_y \binom{n}{y}\theta^y(1-\theta)^{n-y} y\right) \quad (A2.5)$$

It is a standard result for the binomial distribution that $E(y|\theta,n) = n\theta$, hence:

$$h(\theta) = n(1-\theta)^{-2} + (1-\theta)^{-2}(\theta^{-2} - 2\theta^{-1})n\theta \quad (A2.6)$$

which upon simplification becomes

$$h(\theta) = n(1+\theta^{-1}-2)(1-\theta)^{-2} = n\theta^{-1}(1-\theta) \quad (A2.7)$$

Jeffreys' noninformative prior is $\pi(\theta) = |\det(h(\theta))|^{1/2}$, so

$$\pi(\theta) \propto \theta^{-1/2}(1-\theta)^{-1/2} \quad (A2.8)$$

*b) Experiment B: Fixed number of successes – Negative binomial (Pascal) distribution*

Let the probability of success in any trial be $\theta$, the specified number of successes be $r$, and the number of trials that are required to achieve the specified number of successes be $z$. Then:

$$P(z|\theta,r) = \binom{z-1}{r-1}\theta^r(1-\theta)^{z-r} \quad (A2.9)$$

$$\log P(z|\theta,r) = \log\binom{z-1}{r-1} + r\log\theta + (z-r)\log(1-\theta) \quad (A2.10)$$

$$\frac{\partial^2}{\partial\theta^2}\log P(z|\theta) = -r\theta^{-2} - (z-r)(1-\theta)^{-2} \quad (A2.11)$$

Fisher's information (matrix) $h(\theta)$ is:

$$h(\theta) = -E\left(P(z|\theta,r)\frac{\partial^2}{\partial\theta^2}\log P(z|\theta,r)\right) = \sum_z \binom{z-1}{r-1}\theta^r(1-\theta)^{z-r}(r\theta^{-2}+(z-r)(1-\theta)^{-2}) \quad (A2.12)$$

which since $\sum_z P(z|\theta,r) = 1$ simplifies to:

$$h(\theta) = r\theta^{-2} + (1-\theta)^{-2}\left(\sum_z \binom{z-1}{r-1}\theta^r(1-\theta)^{z-r}(z-r)\right) \quad (A2.13)$$

It is a standard result for the negative binomial distribution that $E(z| \theta,r) = r / \theta$, hence:

$$h(\theta) = r\theta^{-2} + (1-\theta)^{-2}(r/\theta - r) \qquad (A2.14)$$

which upon simplification becomes

$$h(\theta) = r\theta^{-2}(1-\theta)^{-1} \qquad (A2.15)$$

Jeffreys' noninformative prior is $\pi(\theta) = |\det(h(\theta))|^{1/2}$, so

$$\pi(\theta) \propto \theta^{-1}(1-\theta)^{-1/2} \qquad (A2.16)$$

c) Combination of experiments A and B – Binomial and negative binomial distributions

Defining the variables as in cases a) and b), we have:

$$P(y,z \mid n,r,\theta) = \binom{n}{y}\theta^y(1-\theta)^{n-y}\binom{\theta-1}{r-1}\theta^r(1-\theta)^{z-r} \qquad (A2.17)$$

$$\log P(y,z \mid \theta,n,r) = \log\binom{n}{y} + y\log\theta + (n-y)\log(1-\theta) + \log\binom{\theta-1}{r-1} + r\log\theta + (z-r)\log(1-\theta) \qquad (A2.18)$$

$$\frac{\partial^2}{\partial\theta^2}\log P(y,z \mid \theta,n,r) = -(y+r)\theta^{-2} - (n-y+z-r)(1-\theta)^{-2} \qquad (A2.19)$$

$$h(\theta) = -E\left(P(y,z \mid \theta,n,r)\frac{\partial^2}{\partial\theta^2}\log P(y,z \mid \theta,n,r)\right)$$
$$= \sum_y \sum_z \binom{n}{y}\theta^y(1-\theta)^{n-y}\binom{\theta-1}{r-1}\theta^r(1-\theta)^{z-r}((y+r)\theta^{-2} + (n-y+z-r)(1-\theta)^{-2}) \qquad (A2.20)$$

which since $\sum_y P(y \mid \theta,n) = 1$, $\sum_z P(z \mid \theta,r) = 1$, $E(y|\theta,n) = n\theta$ and $E(z| \theta,r) = r/\theta$ simplifies to:

$$h(\theta) = r\theta^{-2} + (n-r)(1-\theta)^{-2} + n\theta(\theta^{-2} - (1-\theta)^{-2}) + (r/\theta)(1-\theta)^{-2} \qquad (A2.21)$$

$$h(\theta) = \theta^{-2}(1-\theta)^{-2}\{r(1-2\theta+\theta^2) + (n-r)\theta^2 + n\theta((1-2\theta+\theta^2) - \theta^2) + (r\theta)\} \qquad (A2.22)$$

$$h(\theta) = \theta^{-2}(1-\theta)^{-2}\{r(1-\theta) + n\theta(1-\theta)\} \qquad (A2.23)$$

$$h(\theta) = (r+n\theta)\theta^{-2}(1-\theta)^{-1} \qquad (A2.24)$$

Jeffreys' noninformative prior for the experiments combined is therefore:

$$\pi(\theta) \propto (r+n\theta)^{1/2}\theta^{-1}(1-\theta)^{-1/2} \qquad (A2.25)$$

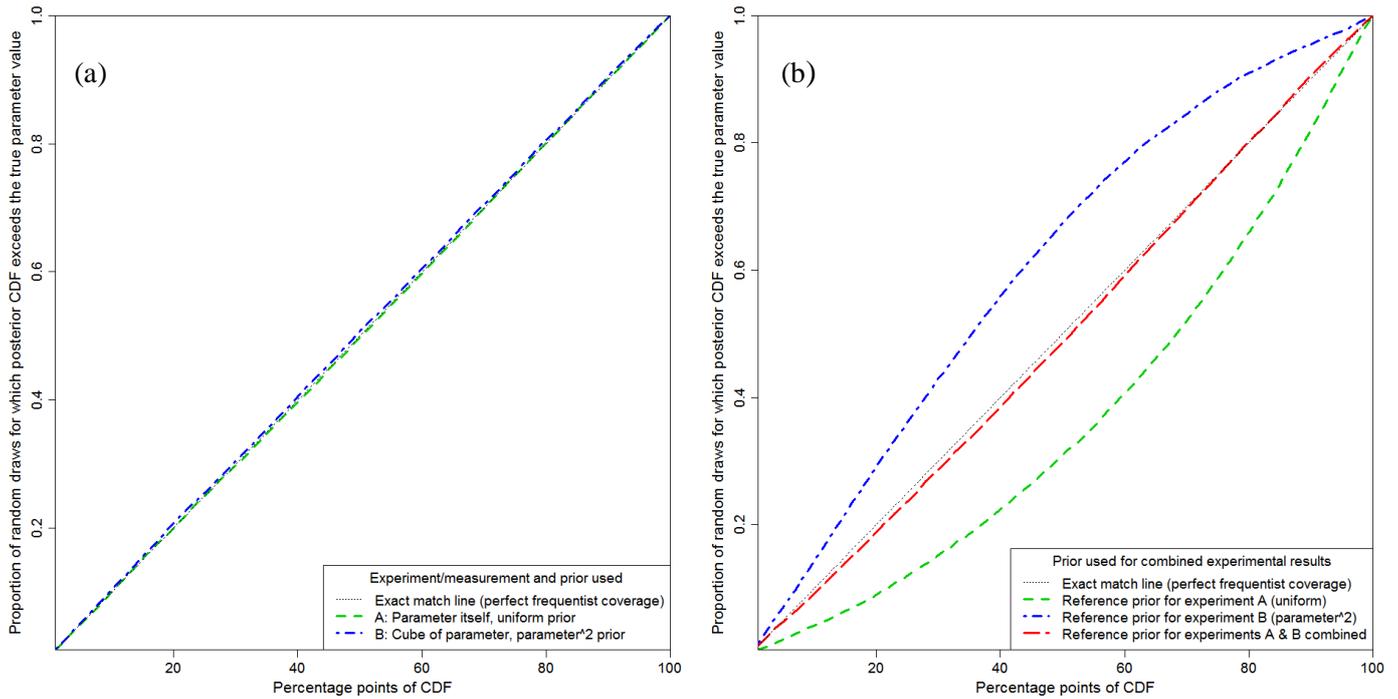

Fig 1: Coverage probabilities – Example 1. Experiments measure: A – parameter; B – cube of parameter.

x-axis: 0–100 percentage points of the posterior CDF for the parameter.

y-axis: proportion of 20,000 randomly drawn experimental measurements for which the parameter value at each posterior CDF percentage point exceeds the true parameter value.

Panel (a): inference based on data from each separate experiment, using the correct Jeffreys'/reference prior for that experiment.

Panel (b): inference based on combined experiments, using the Jeffreys'/reference prior for experiment A (green line), that for experiment B (blue line) or that for the combined experiments (red line).

Black dotted lines show perfect frequentist coverage (both panels).

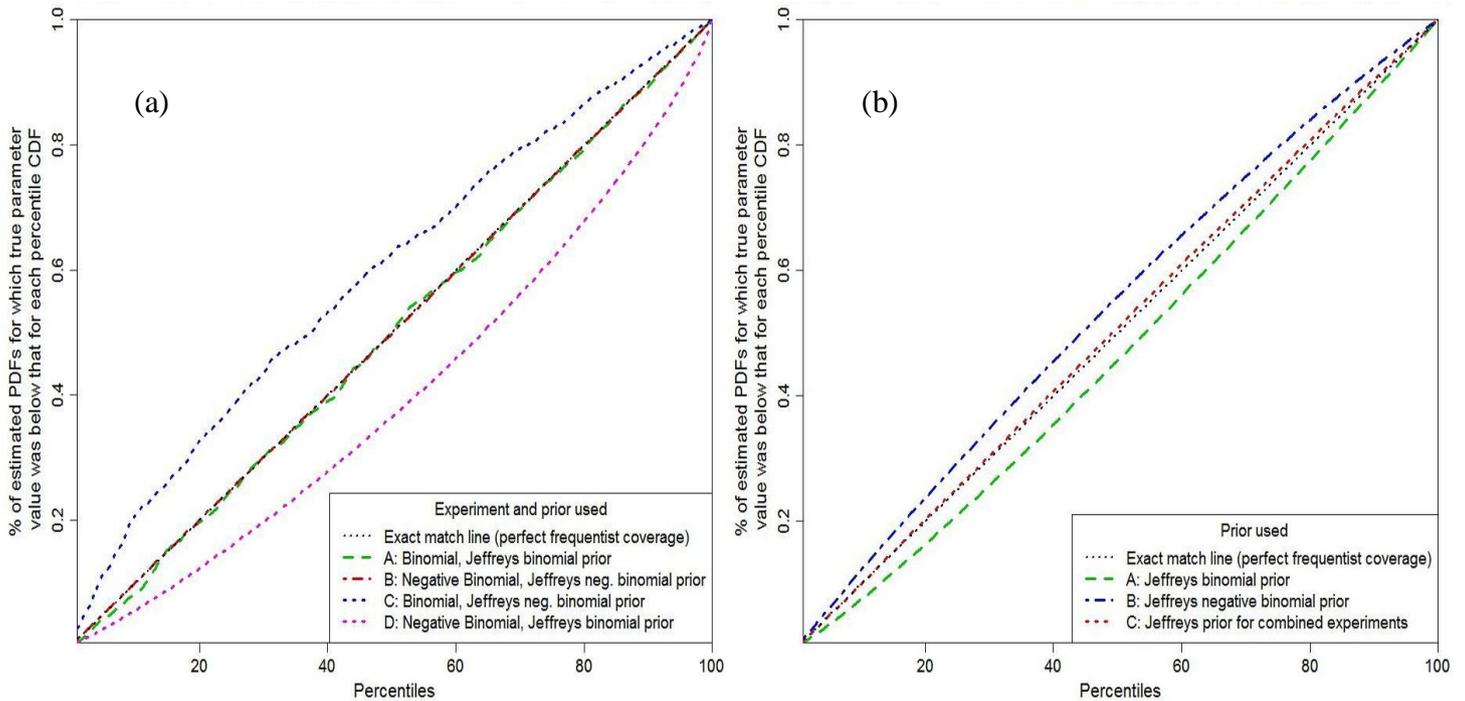

Fig 2: Coverage probabilities – Example 2. Bernoulli experiments: A – binomial; B – negative binomial.

x-axis: 0–100 percentage points of the posterior CDF for the parameter.

y-axis: proportion of 2,000 randomly drawn experimental measurements at each of 100 parameter values randomly drawn between 0.89 and 0.99 for which the parameter value at each posterior CDF percentage point exceeds the true parameter value.

Panel (a): inference based on data from each separate experiment. Green dashed and red dashed-dotted lines are using the Jeffreys' prior appropriate to the relevant experiment, being respectively a binomial and a negative binomial experiment. Blue and magenta short-dashed lines are for the same experiments but with the priors swapped.

Panel (b): inference based on combined experiments, using the Jeffreys' prior for experiment A (green dashed line), that for experiment B (dash-dotted blue line) or that for the combined experiments (red short-dashed line). Black dotted lines show perfect frequentist coverage (both panels).